\documentclass[preprint]{aastex}
\usepackage{natbib}
\newcommand{\blast}{BLASTPol}
\newcommand{\two}{$250\:\mu$m}
\newcommand{\three}{$350\:\mu$m}
\newcommand{\five}{$500\:\mu$m}

\title{Lupus I Observations from the 2010 Flight of the Balloon-borne Large Aperture Submillimeter Telescope for Polarimetry}

\author{Tristan G.\ Matthews\altaffilmark{1},
Peter A. R. Ade\altaffilmark{2},
Francesco E. Angil\`e\altaffilmark{3},
Steven J. Benton\altaffilmark{4},
Edward L. Chapin\altaffilmark{5}, 
Nicholas L. Chapman\altaffilmark{1},
Mark J. Devlin\altaffilmark{3}, 
Laura M. Fissel\altaffilmark{4},
Yasuo Fukui\altaffilmark{6},
Natalie N. Gandilo\altaffilmark{4},
Joshua O. Gundersen\altaffilmark{7},
Peter C. Hargrave\altaffilmark{2},
Jeffrey Klein\altaffilmark{3},
Andrei L. Korotkov\altaffilmark{8},
Lorenzo Moncelsi\altaffilmark{9},
Tony K. Mroczkowski\altaffilmark{9},
Calvin B. Netterfield\altaffilmark{4},
Giles Novak\altaffilmark{1},
David Nutter\altaffilmark{2},
Luca Olmi\altaffilmark{10,11},
Enzo Pascale\altaffilmark{2},
Fr{\'e}d{\'e}rick Poidevin\altaffilmark{12},
Giorgio Savini\altaffilmark{12},
Douglas Scott\altaffilmark{13},
Jamil A. Shariff\altaffilmark{4},
Juan Diego Soler\altaffilmark{4},
Kengo Tachihara\altaffilmark{6},
Nicholas E. Thomas\altaffilmark{7},
Matthew D. P. Truch\altaffilmark{3},
Carole E. Tucker\altaffilmark{2},
Gregory S. Tucker\altaffilmark{8},
Derek Ward-Thompson\altaffilmark{14}}

\altaffiltext{1}{Center for Interdisciplinary Exploration and Research in Astrophysics (CIERA) and Department\ of Physics \& Astronomy, Northwestern University, 2145 Sheridan Road, Evanston, IL 60208}
\altaffiltext{2}{Cardiff University, School of Physics \& Astronomy, Queens Buildings, The Parade, Cardiff, CF24 3AA, U.K.} 
\altaffiltext{3}{Department of Physics \& Astronomy, University of Pennsylvania, 209 South 33rd Street, Philadelphia, PA, 19104, U.S.A.} 
\altaffiltext{4}{Department of Astronomy \& Astrophysics, University of Toronto, 50 St. George Street Toronto, ON M5S 3H4, Canada}
\altaffiltext{5}{XMM SOC, ESAC, Apartado 78, 28691 Villanueva de la Ca\~nada, Madrid, Spain}
\altaffiltext{6}{Department of Physics, Nagoya University, Nagoya 464-8602, Japan}
\altaffiltext{7}{Department of Physics, University of Miami, 1320 Campo Sano Drive, Coral Gables, FL, 33146, U.S.A.}
\altaffiltext{8}{Department of Physics, Brown University, 182 Hope Street, Providence, RI, 02912, U.S.A.}
\altaffiltext{9}{California Institute of Technology, 1200 E. California Blvd., Pasadena, CA, 91125, U.S.A.}
\altaffiltext{10}{University of Puerto Rico, Rio Piedras Campus, Physics Dept., Box 23343, UPR station, San Juan, Puerto Rico}
\altaffiltext{11}{Osservatorio Astrofisico di Arcetri, INAF, Largo E. Fermi 5, I-50125, Firenze, Italy}
\altaffiltext{12}{Department of Physics \& Astronomy, University College London, Gower Street, London, WC1E 6BT, U.K.}
\altaffiltext{13}{Department of Physics \& Astronomy, University of British Columbia, 6224 Agricultural Road, Vancouver, BC V6T 1Z1, Canada}
\altaffiltext{14}{Jeremiah Horrocks Institute, University of Central Lancashire, PR1 2HE, U.K.}

\begin{document}

\begin{abstract}

The Balloon-borne Large Aperture Submillimeter Telescope for Polarimetry (BLASTPol) was created by adding polarimetric capability to the BLAST experiment that was flown in 2003, 2005, and 2006.  BLASTPol inherited BLAST's 1.8 m primary and its \textit{Herschel}/SPIRE heritage focal plane that allows simultaneous observation at 250, 350, and \five.  We flew BLASTPol in 2010 and again in 2012.  Both were long duration Antarctic flights.  Here we present polarimetry of the nearby filamentary dark cloud Lupus I obtained during the 2010 flight.  Despite limitations imposed by the effects of a damaged optical component, we were able to clearly detect submillimeter polarization on degree scales.  We compare the resulting BLASTPol magnetic field map with a similar map made via optical polarimetry (The optical data were published in 1998 by J. Rizzo and collaborators.). The two maps partially overlap and are reasonably consistent with one another. We compare these magnetic field maps to the orientations of filaments in Lupus I, and we find that the dominant filament in the cloud is approximately perpendicular to the large-scale field, while secondary filaments appear to run parallel to the magnetic fields in their vicinities.  This is similar to what is observed in Serpens South via near-IR polarimetry, and consistent with what is seen in MHD simulations by F. Nakamura and Z. Li.

\end{abstract}

\doublespace

\section{Introduction}

The Galactic star formation rate and initial mass function are reasonably well established, but the physical mechanisms that control the initial stages of star formation and determine these observables are not well constrained \citep{mckee2007theory}. Simple models of quiescent gas collapsing under self-gravity give star formation rates that are at least an order of magnitude higher than the observed rate \citep[][and references therein]{netterfield2009blast} requiring the existence of some additional factor to slow the rate of star formation. The two leading theories for this additional support are a strong magnetic field preventing collapse, which is eventually overcome by ambipolar diffusion \citep{mouschovias1999magnetic}, or strong turbulent forces, which can dissipate clouds before they are able to collapse \citep{mac2004control}. It is of course possible that both effects play significant roles with neither fully dominating.

Much work has been done to measure the environments of star forming regions, but observing magnetic fields is experimentally challenging \citep{crutcher2012magnetic}. Measurements of Zeeman splitting of molecular lines provide the most direct probe of magnetic fields in star forming clouds, but this technique is observationally difficult due to Doppler broadening. Consequently there are relatively few clear detections of Zeeman splitting of molecular lines. It is theoretically possible to reconstruct the full 3-dimensional magnetic field with Zeeman measurements, but due to experimental difficulties usually only the line of sight component of the field is measured.

It is also possible to probe the magnetic field by taking advantage of the tendency for spinning dust grains to align with their long axes preferentially perpendicular to the local magnetic field \citep{lazarian2007tracing}. This alignment leads to a slight polarization in light that interacts with the dust. Transmitted light from background stars is preferentially absorbed by the long axis of the grain causing a slight polarization along the local field direction. In addition, dust grains preferentially emit light polarized parallel to their long axes causing a net polarization perpendicular to the local magnetic field. Because interstellar dust grains have temperatures of $T\approx10-30$ K, they emit most strongly at sub-millimeter (submm) wavelengths. Polarization pseudo-vectors are measured in the plane of the sky giving the inferred direction of the projected magnetic field. These dust polarization studies, however, provide no direct measure of magnetic field strength. While still experimentally challenging, dust polarization mapping is possible for a wide range of clouds with different cloud conditions. Investigators have used optical polarimetry to map the outer diffuse regions of molecular clouds, but this technique cannot probe the highly extincted inner regions \citep{heiles20009286}. More recently, submm studies have been able to map the denser central regions of clouds providing a clearer view into the regions where star formation is actually occurring \citep{matthews2009legacy,dotson2010350}.

Field strength measurements derived from Zeeman observations typically imply mass-to-magnetic flux ratios that are supercritical by a factor of $\sim$2-3, meaning the field is not strong enough to support the cloud against gravitational collapse. However, due to systematic uncertainties, the prevalence of magnetically critical or balanced clouds cannot be ruled out \citep{crutcher2012magnetic}. Less direct methods for probing magnetic field strength have also been employed; \citet{novak2009dispersion} compared the dispersion in inferred magnetic field direction, as measured with submm polarimetry, with that of simulated clouds from \citet{ostriker2001density}. They concluded that agreement between observations and simulations is best when the total magnetic energy density is at least as large as the turbulent kinetic energy density. A number of studies have probed the relationship between observed magnetic field directions and the morphology of cloud substructure.  For example \citet[also see Basu 2000]{ward2000first,ward2009optical} mapped submm and optical polarization toward starless cores and observed that, in projection, their minor axes are offset by about thirty degrees, on average, relative to the magnetic field direction. \citet{tassis2009statistical} found a similar result on larger scales, via statistical analysis of submm polarimetric observations toward 24 sources.  They modeled cloud substructure using a simple spheroidal model, finding a preference for oblate versus prolate shape and a positive correlation between minor axis and magnetic field.  Such a correlation seems inconsistent with some turbulent models; e.g., the model of \citet{gammie2003analysis} predicts a random relationship between cloud substructure and field direction. \citet{goodman1990optical} investigated the magnetic morphology of cloud complexes by observing filamentary structures in Taurus, Ophiuchus, and Perseus via optical polarimetry, and found that the magnetic field direction shows little variation over $\geq10$ pc size scales, while individual filamentary clouds within a complex show significant variation in their orientation with respect to the field direction. Thus, the magnetic field appears to be strong enough to cause large-scale correlations and significantly affect the morphology of cloud substructure, but not so strong that it defines all features of molecular clouds.

The Lupus I molecular cloud has a number of properties that make it ideal for studying the early stages of star formation. At a distance of ($155\pm8$) pc \citep{lombardi2008hipparcos}, it is one of the closest star-forming regions. It is also well separated from the Galactic plane (b $\sim16^\circ$) with little confusion along its line of sight. Finally, while \textit{Herschel} has observed numerous prestellar and young stellar objects \citep{rygl2013},  the cloud is not disturbed by the presence of O and B stars which more violently disrupt their environment.

Here we report on submm polarimetric measurements of Lupus I made during the 2010 flight of the Balloon-borne Large Aperture Submillimeter Telescope for Polarimetry (\blast). \blast{} incorporates a 1.8 m parabolic primary mirror and large-format bolometer arrays operating at \two, \three, and \five{} simultaneously. Coded photo-etched masks of vertical and horizontal polarizing grids are placed in front of the array feed horns to provide polarization sensitivity. \blast's predecessor, BLAST, was a total intensity experiment that was successfully flown in 2003, 2005, and 2006, with results reported in many publications \citep[e.g.,][]{devlin2009over,netterfield2009blast}.

\section{\label{sec:obs}Observations}

A full description of BLAST can be found in \citet{pascale2008balloon}. For \blast, we modulate the polarization in two ways. First, we use a rotating achromatic half-wave plate \citep{moncelsi2012empirical} in the optical path. Secondly, the coded mask of grids for each array alternates between vertical and horizontal wire directions as one moves along the array in the horizontal (scan) direction, thus providing for rapid modulation of a given source's polarization during spatial scanning of the gondola. A detailed description of the upgrade of BLAST into \blast{} is provided by Angil\`e et al. (2013, in preparation) who also summarize all observations taken during the two Antarctic flights of \blast, which took place in 2010 and 2012. The present paper deals only with Lupus I results from the 2010 flight.

Our 2010 flight began with a launch from McMurdo Station, Antarctica, on December 27th, 2010, and continued for 9.5-days. We made $\approx49$ hours of scan mode observations on the Lupus I molecular cloud, resulting in 0.91, 1.02, and 1.06 deg$^2$ maps at 500 and 350, and \two{} respectively. The target region was covered with two passes (up and down in elevation) while continuously scanning in azimuth before stepping the Half-wave Plate (HWP) and repeating the process. A `data chunk' consists of two such passes at a single HWP position. Four HWP positions were used, separated from each other by $22.5^\circ$. Scan speeds were chosen so that the HWP was stepped approximately every 15 minutes, and observations were generally made at all four HWP positions before changing targets.

Preflight calibration showed low instrumental polarization (IP $\approx0.2\%$) and nearly Gaussian, diffraction limited beams at all three wavelengths. Unfortunately, during ascent to observing altitude, the outermost IR blocking filter became partly melted, which introduced significant systematic error. As discussed in Section \ref{sec:data_analysis}, the in-flight measurements of instrumental polarization were dramatically higher than pre-flight calibration, at 5.8\%, 6.0\% and 6.8\% for \five, \three{} and \two, respectively. Additionally, there was a significant degradation in the beam shape, which resulted in increased beam size with non-Gaussian structure and polarized point spread functions. To mitigate this, all three wavelengths were smoothed using a $2\farcm5$ FWHM Gaussian kernel. A grid of polarization measurements were then calculated, with grid spacing also equal to $2\farcm5$. We believe the analysis pipeline that we summarize in Section \ref{sec:data_analysis} has been able to remove or account for the above-mentioned systematic errors sufficiently well to allow us to report a meaningful measurement of large-scale polarization structure in the Lupus I cloud. Preliminary analysis of observations from our second Antarctic flight (the 2012 flight) indicates low levels of in-flight instrumental polarization consistent with ground tests, i.e., under 1\%. Polarimetry results from the 2012 flight are under analysis and will be reported at a later time.

\section{\label{sec:data_analysis}Data Analysis}

The \blast{} data analysis pipeline can be broken down into six basic parts, which are described, respectively, in Sections 3.1 through 3.6. A more detailed discussion of the data analysis pipeline will be presented by Fissel et al. (2013, in preparation).

\subsection{Time Ordered Data Processing}

We first employed standard preprocessing techniques to de-spike, i.e., to remove cosmic rays, and to deconvolve the detector bolometer response functions. In order to remove a significant elevation-dependent feature from the raw bolometer Time Ordered Data (TOD), we fit to an elevation dependent model for each data chunk. During the flight an unusual noise feature was recognized in the TOD series. It was characterized by rapid changes in the DC level of otherwise normal data. It was later determined that this was due to a combination of motor current cross talk from the fly wheel motor and pick up from the high gain Tracking and Data Relay Satellite System (TDRSS) antenna. Both problems were intermittent. The correlation between each channel and a dark channel of the same wavelength was calculated for each data chunk, data chunks with high correlation were flagged and are not used in the analysis.

Preprocessed and flagged TOD series were passed to the naive mapmaker `naivepol' \citep{moncelsi2012empirical}, which generates $I$, $Q$ and $U$ Stokes parameter maps via a binning algorithm. To remove 1/f noise, a 0.05 Hz high-pass filter was applied to the TOD series before binning. Observations were binned into 20'' map pixels. As noted earlier, the maps were later Gaussian smoothed with a $2\farcm5$ kernel.

\subsection{\label{sec:characterization_ip}Characterization of Instrumental Polarization}

The instrumental polarization (IP) was characterized by taking advantage of the fact that celestial sources rotate relative to our alt-az telescope as they rise and set, while instrumental polarization remains unchanged \citep{hildebrand2000primer}. To maximize signal-to-noise ratio, the luminous source Vela C was used for this IP calibration. Observations were broken into two bins based on the parallactic angle of the source. A synthetic aperture was defined around bright regions of the intensity map, with two low flux regions chosen for reference. Mean polarization parameters were then calculated for each detector for both sky rotation bins. These measurements were next fit for source polarization and individual bolometer instrumental polarization, holding the mean sky rotation difference of $8.5^\circ$ between the two bins as a fixed parameter. The individual bolometer instrumental polarization terms were then passed to naivepol, which was modified to subtract out the instrumental polarization while binning maps.

\subsection{\label{sec:map_referencing}Map Referencing}

Due to the high pass filtering in naivepol, \blast{} is insensitive to large-scale spatial modes. This causes our maps to have zero mean so that large regions of the map have negative flux. Polarization measurements are usually specified as percent polarization, which diverges at an intensity of zero. To account for this, we computed differential measurements between the $I$, $Q$, and $U$ Stokes parameters corresponding to the high intensity points of interest and the mean Stokes parameters of the large reference regions shown in Figure \ref{fig:spire}. Reference regions were chosen using the \textit{Herschel} SPIRE \three{} intensity map, which contains no significant filtering artifacts. The choice of reference regions was tested with simulations (see Section \ref{sec:simulations}).

This `reference region' technique can be thought of as fixing the zero spatial mode to the mean value of the reference regions which allows \blast{} polarization pseudo-vector maps to be studied in a traditional manner \citep[see][]{hildebrand2000primer}. Our technique does not make any attempt to account for other lost spatial modes. Accordingly, an intensity cut was applied requiring all reported pseudo-vectors be in regions of the map having positive unreferenced flux. To fully reconstruct all spatial modes will require the development of a new mapmaker, which is beyond the scope of the present paper.

\subsection{Null Tests}

Due to the large systematic errors introduced by the melted filter, statistical error bars do not provide a meaningful characterization of the validity of the data. Instead, a series of null tests were employed to gauge which measurements were robust and repeatable. For each test, the data was divided into two equal groups (a and b) from which maps were generated. We then calculated residual maps for each Stokes parameter, where the residual of $Q$ is defined by\footnote{Because only the magnitude of the residual is used in the test, $(a - b)/2$ is equivalent to $(b - a)/2$.}:

\begin{equation} Q_\mathrm{residual} = Q_\mathrm{a} - (Q_\mathrm{a} + Q_\mathrm{b})/2 = (Q_\mathrm{a} - Q_\mathrm{b})/2. \end{equation}

We also calculated the residual for $U$ in the same manner. A pseudo-vector is then judged to pass the null test if the magnitude of the polarized flux in the final map is greater than three times the magnitude of the flux for the corresponding sky position in the residual map\footnote{Due to nonlinearities in the mapmaking process the final map is not equal to the mean that is used in the definition of the residual. Passing the full data set into the mapmaker yielded a slightly more accurate map. In practice, the use of the mean or final map for residual calculation or in the final comparison has no significant effect.}:

\begin{equation} (Q^2 + U^2)^{1/2} > 3 \times (Q_\mathrm{residual}^2 + U_\mathrm{residual}^2)^{1/2}. \end{equation}

We performed six null tests to investigate different potential systematic errors. Three of the tests were temporal, based on when the data was acquired: Early flight / Late flight, High Elevation / Low Elevation, and High Parallactic Angle / Low Parallactic Angle. The other three tests were array cuts, based on which bolometer made the measurements (dividing the array into two equal groups): Left Half / Right Half, Top Half / Bottom Half, and Large IP Correction / Small IP Correction\footnote{As defined by the magnitude of the instrumental polarization correction, which is applied as described in \ref{sec:characterization_ip}.}. Only pseudo-vectors that pass all six null tests are reported.

\subsection{\label{sec:simulations}Data Simulations}
Source polarization and instrumental polarization reconstruction fidelity were tested using simulated \blast{} \five{} observations. Simulated TOD series were generated using the instrument simulation software Simsky, originally developed for BLAST and since adapted for polarimetry. The software `observes' input $I$, $Q$, and $U$ maps convolved with a beam (defined in the telescope reference frame), for given input pointing and HWP angle TOD series. For the simulations discussed here, input Stokes I maps were based on \textit{Herschel} SPIRE observations of Vela\,C \citep{hill2011filaments} and Lupus I \citep{rygl2013}\footnote{http://herschel.esac.esa.int/Science\_Archive.shtml}. A constant polarization percentage was assumed: $Q$=$q$I and $U$=$u$I, where $q$ and $u$ were fixed over the map. Pointing TOD series were taken from the \blast{} 2010 pointing solution scans, HWP angles were also taken from the 2010 data. Noise TOD series were generated with a white noise component for each individual detector based on 2010 measurements, and a correlated noise component common to all bolometers based on the following noise spectrum:
\begin{equation} PSD_{corr}\,=\,W\frac{f_0}{f}, \label{eqn:null_diff} \end{equation}
Here $W$ is the median white noise level for the \five{} bolometers, and $f_0$ is the median 1/f knee.

The simulator also adds polarized point spread functions q$_{IP}^{i}$, u$_{IP}^{i}$ to the TOD series for each bolometer, constructed as follows: The beam shape was assumed to be identical for all detectors with the same grid orientation (i.e., all horizontal detectors have the same beam and all vertical detectors have the same beam). Beam patterns were generated from fits of three independent elongated Gaussians to maps made from the \five{} observations of the nearly point-like source IRAS\,08470-4243 at half-wave plate angles $0^\circ$ and $67.5^\circ$ \footnote{There was insufficient data at half-wave plate angles $22.5^\circ$ and $45^\circ$ to make high signal-to-noise maps, so for the purposes of these simulations it was assumed that the beam at $0^\circ$ is identical to the beam at $22.5^\circ$ and the beam at $45^\circ$ is identical to the beam at $67.5^\circ$.}.

Simsky was used to generate TOD series for both Vela C and Lupus I using the IP and polarized point spread function prescription. Simulated TOD series were passed through the \blast{} pipeline as preprocessed TOD series. Instrumental polarization parameters were calculated on Vela C and applied to Lupus I maps in the same manner as for the science TOD series. Large-scale polarization of Lupus I was reconstructed to within 0.46\%, which provides an estimate of the systematic error level. Pseudo-vectors having degree of polarization below 1.5\% (approximately three times the above error level) are discarded regardless of whether they pass the six null tests. The simulations and data cuts described in this section were carried out prior to application of the polarization efficiency correction (see section \ref{sanity}).

\subsection{\label{sanity}Consistency Checks}

During our 2010 flight, we observed two targets that had previously been observed by the SPARO polarimeter at $450\:\mu$m \citep{li2006results}. One is Carina Nebula, a relatively high polarization source that we observed as a check on the calibration of the polarization zero angle. \blast{} and SPARO observe very similar patterns of angles \citep[see][]{pascale2012balloon}. The \blast{} 2010 Carina Nebula maps are not large enough to include the reference regions used by SPARO so it is impossible to account for fundamental differences in observing strategy. Consequently, ground measurements are used for calibration of polarization zero angles. We believe that these are accurate to $\pm2^\circ$.

The second SPARO target observed is G331.5-0.1, which SPARO found to have low, $\sim$0.3\%, polarization near its intensity peak. Again, \blast{} maps did not include the SPARO reference regions so G331.5-0.1 cannot be used for calibration, but \blast{} \five{} measurements of (q,u) near the intensity peak were found to differ by roughly 0.5\% from SPARO's reported $450\:\mu$m values, providing a crude consistency check on our IP correction. Under-sampling effects due to the loss of significant amounts of \blast{} data to correlated noise while observing this source may account for the $\approx0.5$\% discrepancy.

The polarization efficiencies were determined from ground measurements to be approximately equal to 79\% and 82\% for the $350\:\mu$m and $500\:\mu$m  bands, respectively, and the results were corrected accordingly.  Due to the above-mentioned differences between \blast{} and SPARO reference regions, it was not possible to check these values using SPARO measurements.

\section{\label{sec:results}Results}

\begin{figure} \plotone{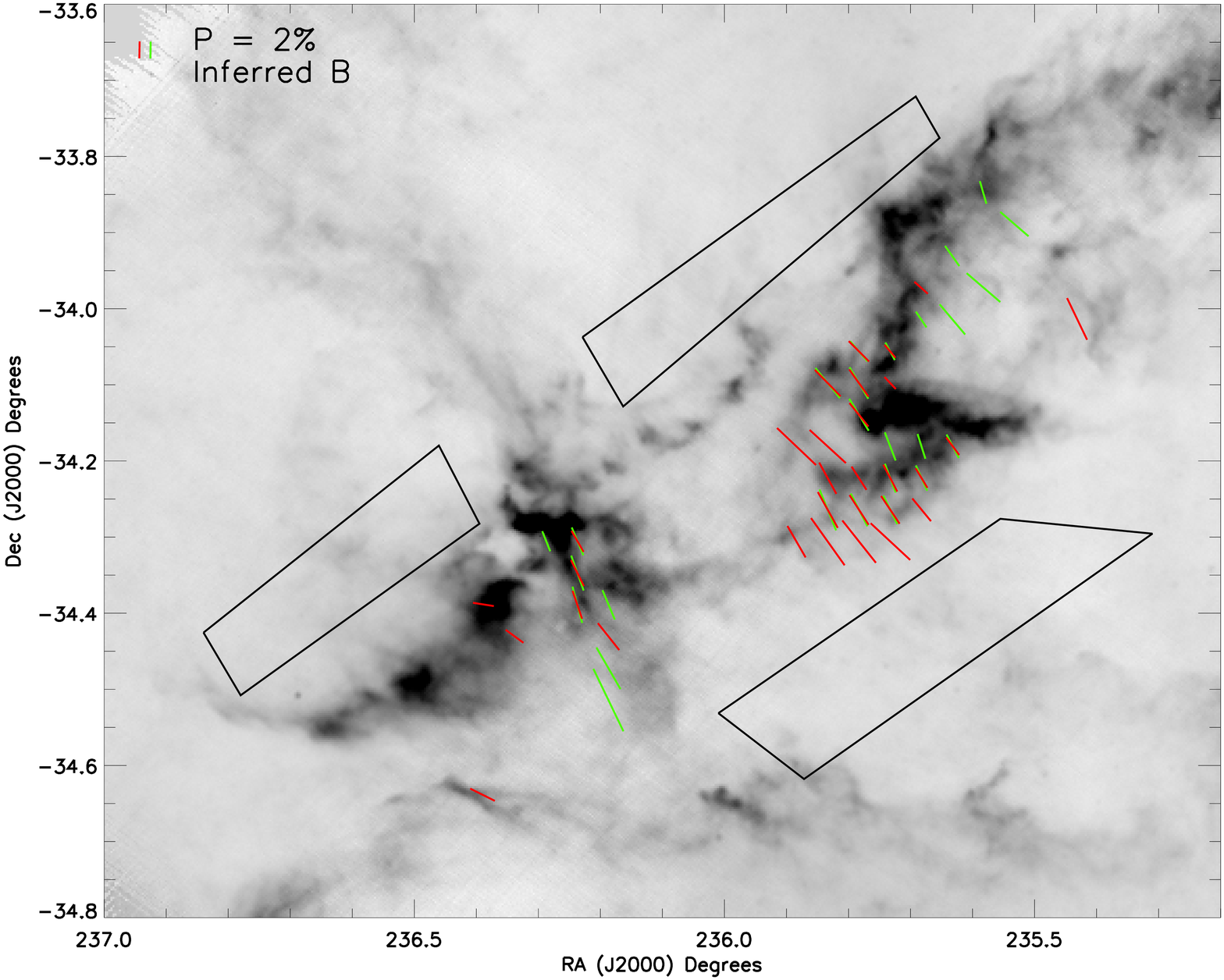}

\caption{\label{fig:spire}\blast{} submm polarimetry superimposed on \textit{Herschel} SPIRE \three{} map of Lupus I (grayscale). Red (green) pseudo-vectors show \five{} (\three) measurements, and the orientation of each pseudo-vector is drawn parallel to the inferred field direction (perpendicular to the orientation of the $E$-vector of the polarized radiation). \blast{} reference regions (outlined in black) are used for differential measurement in deriving submm pseudo-vectors. Pseudo-vector length is proportional to degree of polarization (see key at upper left).}

\end{figure}

\begin{figure} \plotone{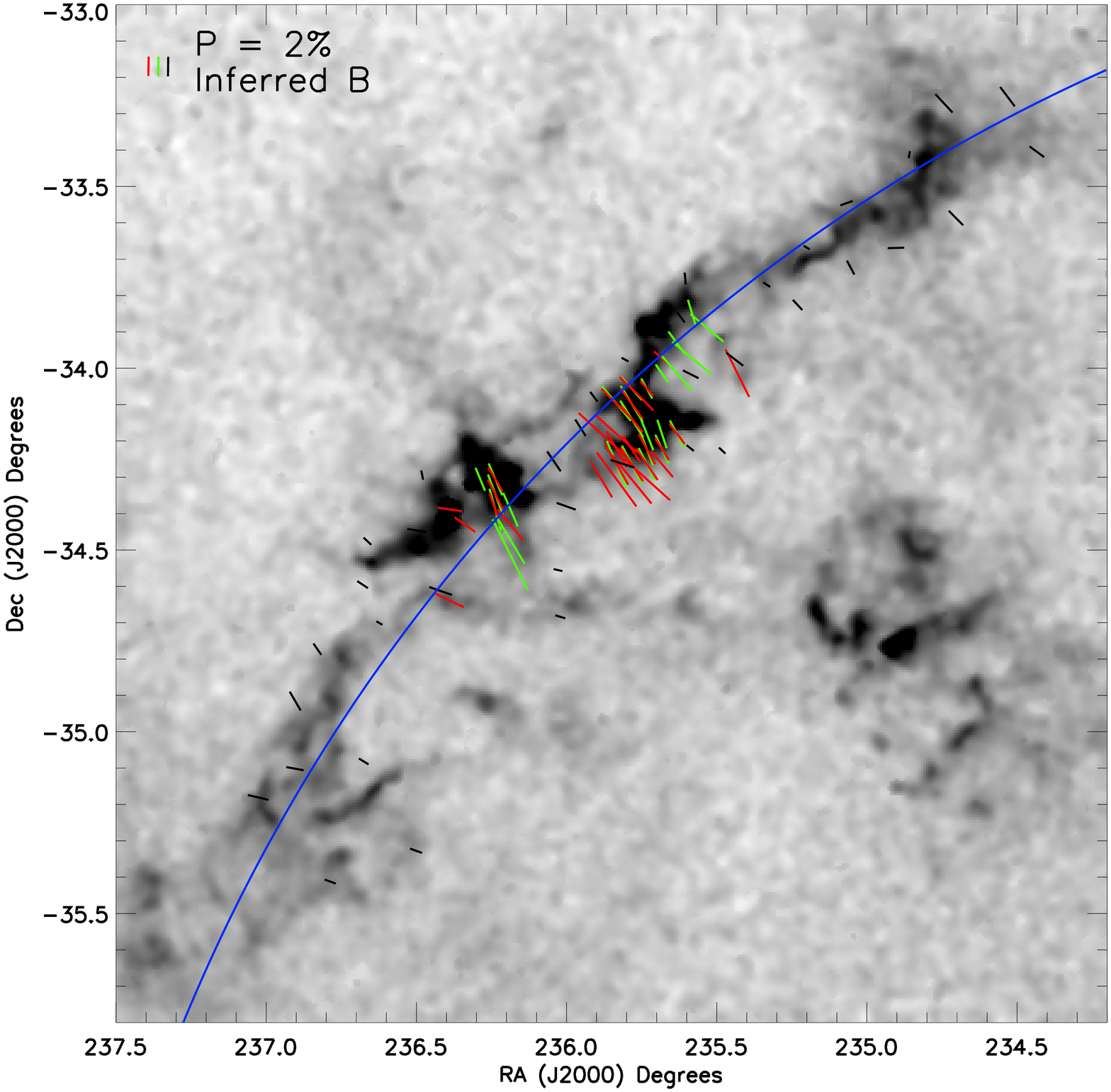}

\caption{\label{fig:avmap} \blast{} submm polarimetry (red and green pseudo-vectors) superimposed on $A_V$ map of Lupus I (grayscale). Red (green) pseudo-vectors show \five{} (\three) measurements, and the orientation of each pseudo-vector is drawn parallel to the inferred field direction (perpendicular to the orientation of the $E$-vector of the polarized radiation). Also shown are optical polarimetry pseudo-vectors from \citet{rizzo1998starlight}, in black. The blue arc represents a `by eye' fit to the main filament, as discussed in Section \ref{sec:magnectic_models}. As in Figure \ref{fig:spire} pseudo-vector length is proportional to degree of polarization.}

\end{figure}

Submillimeter pseudo-vectors that survived all analysis criteria (see Section \ref{sec:data_analysis}) are displayed in Figure \ref{fig:spire} for both \five{} (red) and \three{} (green); no significant detections were obtained at \two. The background image is a portion of the \three{} map from SPIRE \citep{rygl2012recent}\footnote{http://herschel.esac.esa.int/Science\_Archive.shtml}. Also shown are the submm reference regions (see Section \ref{sec:map_referencing}). Figure \ref{fig:avmap} shows a larger view of the region and includes optical polarimetry data (black) from \citet{rizzo1998starlight}, plotted on a $2'$ resolution extinction map that we created using the 2MASS near-IR catalogues and the NICER technique \citep{lombardi2001mapping}. Appendix 1 explains how we determined the coordinates of the optical polarization measurements, as these were not provided by \citet{rizzo1998starlight}. Only optical pseudo-vectors which correspond to $3\sigma$ polarization detections are displayed. As can be seen in these figures, the projected structure of the Lupus I cloud is dominated by a single main filament, which is surrounded by a number of secondary filamentary structures. The blue curve in Figure \ref{fig:avmap} represents a `by eye fit' of a circular arc to this primary filament. Figure \ref{fig:avmap} shows that both optical and submm pseudo-vectors are consistent with a general picture of a relatively well ordered large-scale magnetic field oriented approximately perpendicular to the primary filament. However, it is important to keep in mind that molecular clouds are complicated three-dimensional structures that are observed projected into two dimensions.

A visual inspection of the SPIRE map in Figure \ref{fig:spire} shows that the primary filament runs from $\sim$($235.3^\circ$, $-33.7^\circ$) to $\sim$($236.6^\circ$, $-34.55^\circ$) in (R.A., Decl.), giving a position angle of $\approx127^\circ$ east of north for the filament, or a normal of $\approx37^\circ$. If one instead examines the extinction map shown in Figure \ref{fig:avmap} the filament can be seen to extend further to the south east, with overall endpoints at $\sim$($234.65^\circ$, $-33.35^\circ$) and ($\sim$$237.35^\circ$, $-35.55^\circ$), implying a filament normal at position angle of $\approx45^\circ$. Figure \ref{fig:avmap} clearly shows a significant bend to the filament, which is better parameterized by a circle centered at ($231.77^\circ$, $-37.67^\circ$) with a radius of $4.92^\circ$, as shown by the blue arc.

\section{Discussion}

\subsection{\label{sec:magnectic_models}Magnetic Field Models}

In order to quantitatively compare magnetic field directions with filament orientations, we consider two simple models for the magnetic field structure. The first assumes a spatially uniform magnetic field and is referred to as the Uniform Field model. The second assumes that the magnetic field angle has a constant offset with respect to the local primary filament normal, with the filament represented by the blue arc of Figure \ref{fig:avmap}. This is referred to as the Constant Offset model. Because only the direction of the magnetic field is a free parameter in these models, all polarization detections are normalized to unit polarized flux $(Q^2 + U^2 = 1)$ before fitting to models. For both models, we performed least squares fits to these rescaled Stokes parameters. Equal weight is given to all pseudo-vectors. For the submm data, this is appropriate because the dominant errors are believed to be systematic (see Section \ref{sec:data_analysis}). For the optical data, the dispersion in the polarization angles is much larger than the statistical errors of the measurements, which again implies that the equal weight prescription is appropriate. For the Constant Offset model, the filament normal is derived by referencing to the nearest point on the blue arc of Figure \ref{fig:avmap}. The results of these model fits are presented in Table \ref{tab:results}.

As can be seen in Table \ref{tab:results}, the best-fit Uniform Field submm inferred magnetic field angles are $39.2^\circ$ and $28.9^\circ$ for the \five{} and \three{} respectively, and $34.3^\circ$ if all submm pseudo-vectors are taken together. If we reference these angles to the local primary filament normal of $37^\circ$ (see Section \ref{sec:results}) we obtain a value of $\approx-3^\circ$ when using all submm pseudo-vectors together. The best-fit Constant Offset inferred magnetic field angles are $-5.3^\circ$ and $-15.1^\circ$ for \five{} and \three{}, respectively, and $-9.8^\circ$ when all submm pseudo-vectors are used. The optical data shown in Figure \ref{fig:avmap} correspond to polarization measurements obtained by \citet{rizzo1998starlight} on stars they refer to as Weak Stars, i.e. faint stars seen projected onto the high optical extinction regions of Lupus I. While the sky area studied in the submm differs significantly from that studied in the optical, the respective best-fit Uniform Field model angles are not too different at $34.3^\circ$ and $53.9^\circ$, respectively. These angles bracket the large-scale filament normal of $45^\circ$ (see Section \ref{sec:results}). The best-fit Constant Offset field angles are $-9.8^\circ$ and +$8.6^\circ$ for submm and optical, respectively. Thus, while projection effects prevent us from determining the angle between magnetic field and primary filament with certainty, both submm and optical polarization measurements are consistent with a large-scale field that is predominantly perpendicular to the primary filament, as can be also seen in Figure \ref{fig:avmap}. 

\subsection{\label{sec:compare_optical}Comparing Optical and Submm Polarimetry}

Neither the optical nor submm pseudo-vectors uniformly sample the entire primary filament. Rather, they sample spatially distinct regions of the cloud. If there is an ordered large-scale magnetic field perpendicular to the primary filament, then the region of filament sampled should not be critical. If only optical pseudo-vectors within $15'$ of a submm pseudo-vector are considered, best-fit angles for Uniform Field and Constant Offset models of $53.1^\circ$ and $9.0^\circ$ respectively are obtained (see Table \ref{tab:results}), shifts of $-0.8^\circ$ and $0.4^\circ$ respectively. These shifts are not significant, suggesting that the sampling location is not the dominant cause of the difference in angles between the optical and submm data sets. However, while the optical data are relatively evenly spread over the entire filament, the majority of the submm detections are concentrated in a single dense cluster. Any variations in magnetic field direction on the scale of this cluster will be over represented in the submm data, and thus may account for most of the difference in best-fit field angles between the two data sets. Several other factors which might play significant roles in explaining this difference are considered next.

Optical and submm polarimetry are subject to very different experimental biases. \blast{} data were smoothed with a $2\farcm5$ Gaussian beam, thus averaging out any substructure, whereas the optical data are measured on single stars giving `pencil beam' measurements in absorption through regions of the cloud having $A_V$ $\approx2-3$ \citep{rizzo1998starlight}. The optical data are incapable of probing the densest regions of Lupus I, where $A_V$ increases further. The high extinction regions within a submm beam dominate the measurements, while optical polarimetry is only possible at the edges of these regions, which may lead to a sampling bias towards low extinction holes in high extinction regions. Thus, even co-pointed beams might be measuring very different dust populations unless the cloud is uniform on $2\farcm5$ scales, which the SPIRE maps indicate is rarely if ever the case (see Figure \ref{fig:spire}).

Another important difference relates to the cloud environment. \blast{} uses reference regions to compute differential measurements between the dense interiors of the clouds and the diffuse outer regions, making it insensitive to large-scale foreground and/or background contamination. The optical measurements, in contrast, perform no differencing and are measuring the entire column between the observer and star. In addition to their Weak Stars, \citet{rizzo1998starlight} present polarization measurements of Field Stars, which are located within an approximately $6^\circ$ by $8^\circ$ region containing the Lupus I filament. These stars show a very different magnetic field structure than the filament, and many show low degrees of polarization. Using only $3\sigma$ polarization detections, a best-fit Uniform Field direction of $129.3^\circ$ is obtained, showing a roughly $75^\circ$ rotation in the field direction compared to the Weak Stars. To address foreground contamination, we make the simple approximation of a uniform-field foreground medium and estimate its polarization properties from the mean Stokes $Q$ and $U$ of the Field Stars. Then, we subtract this contamination from the Weak Star polarization parameters to obtain corrected angles of $50.1^\circ$ and $5.1^\circ$ for the best-fit Uniform Field and Constant Offset models using all optical pseudo-vectors (see Table \ref{tab:results}), corresponding to relatively small shifts of $3.8^\circ$ and $3.6^\circ$, respectively. However, since none of the Field Stars are in the \blast{} reference regions it is impossible to reference the two data sets in the same manner. This makes it difficult to draw firm conclusions about the precise magnitude of this effect.

Despite the above-mentioned differences between the submm and optical polarimetry techniques and the differences in spatial sampling, both data sets paint a consistent picture of a large-scale magnetic field in the Lupus I cloud that is primarily perpendicular to the primary filament. Previous comparisons of submm and optical polarimetry on the scale of both cloud complexes \citep{li2009anchoring} and Bok globules \citep{ward2009optical} have also shown rough agreement in field directions between these two techniques.

\subsection{The Bipolar Molecular Outflow in B228}

Outflows from protostars are ubiquitous. Theoretical models predict that they arise from circumstellar disks, with the outflow axis parallel to the rotation axis of the disk \citep[e.g.,][]{konigl00,shu00}. Dynamically important magnetic fields may be expected to cause circumstellar disks to form with their rotation axes parallel to the local cloud magnetic field direction due to magnetic braking \citep{mouschovias79}. For this reason, previous authors have looked for a correlation between outflow axis and magnetic field direction, with mixed results \citep[e.g.,][]{menard04,davidson11,targon11,chapman2013alignment,hull13}. The observed area of Lupus I contains a single known molecular outflow, associated with the class 0 protostar B228 \citep[IRAS15398-3359;][]{shirley00}. The outflow is designated HH185, and was discovered in the $K$ band by \citet{heyer89}, who measured it to have a total extent of $15''$ and a position angle of $65^\circ$ east of north. This angle differs from the angles of the submm and optical Uniform Field models by $\sim$$31^\circ$ and $\sim$$11^\circ$, respectively (see Table \ref{tab:results}).

As discussed in Section \ref{sec:compare_optical}, the submm pseudo-vectors will trace the magnetic field in relatively denser regions, in comparison with optical pseudo-vectors. If magnetic braking operates in the manner suggested by \citet{mouschovias79}, we would expect the outflow axis to align with the magnetic field as measured by \blast. Since we find a $\sim$$30^\circ$ difference in angle between the two, it might appear that magnetic braking does not operate in this manner \citep[e.g.,][]{joos12}. However, the inclination angle of the HH185 outflow with respect to the plane of the sky is unknown, and \citet{chapman2013alignment} point out that in this situation we cannot draw strong conclusions based solely on the misalignment angle between the projected magnetic field direction and the projected outflow axis. Magnetic field and outflow directions are also often compared with the elongations of pre-stellar and/or protostellar cores \citep[e.g.][]{ward2000first}. For the case of Lupus I, such a comparison will be presented in a separate paper by Poidevin et al. (2013, in preparation).

\subsection{Secondary Filaments}

We noted earlier that besides the main filament in Lupus I, there is significant substructure in secondary filaments. Figure \ref{fig:spire} shows a secondary filament that runs approximately perpendicular to the main filament, extending both north and south of it, with endpoints near ($236.7^\circ$, $-33.8^\circ$) and ($236.1^\circ$, $-34.6^\circ$). Consider the region of this secondary filament that is sampled by the submm pseudo-vectors just south of the main filament. The submm polarimetry results for this region suggest that the magnetic field direction in the main filament continues into this secondary filament, implying that the magnetic field runs parallel to this secondary filament and not perpendicular to it as in the main filament. Another secondary filament runs at a near constant Decl. of $\approx-34.7^\circ$, and is sampled by several optical pseudo-vectors and a single \five{} pseudo-vector (see Figure \ref{fig:avmap}), all of which show a local field direction running approximately parallel to this secondary filament. Both of these secondary filaments are marginally sampled; it would be unwise to over interpret the data, but the overall magnetic field structure of Lupus I consistent with the idea of a single dominant filament running perpendicular to the mean magnetic field, surrounded by secondary filaments that run more nearly parallel to their local field directions. 

\subsection{Comparison with Other Work}

\citet{nakamura2008magnetically} generated three-dimensional MHD simulations of star formation in turbulent, magnetized clouds, including ambipolar diffusion and feedback from protostellar outflows. The field strength value used in the simulations is high enough to guide the gravitationally driven flow, resulting in formation of sheets or filaments perpendicular to the mean field direction with significant structure in diffuse secondary filaments running approximately parallel to the local field direction. This morphology is consistent with what we have reported for Lupus I. Near-infrared polarimetry of Serpens South \citep{sugitani2011near} also traces a large-scale magnetic field perpendicular to a dominant filament, again surrounded by numerous `sub-filaments' running more parallel to the local field. A similar structure is also observed in \textit{Herschel} SPIRE maps of the B211/3 filament in Taurus as presented \citet{palmeirim2013herschel}. These authors observe low density secondary filaments extending perpendicularly out from the main filament. This main filament is predominantly perpendicular to its local magnetic field as measured by optical and near-IR polarimetry. The secondary filaments are reminiscent of the CO striations observed by \citet{goldsmith2008large} in other low density regions of Taurus, and these CO striations closely follow their local magnetic fields as given by optical polarimetry. Thus, observations of Serpens South, B211/3, and Lupus I, taken together, appear to give strong support to the model of \citet{nakamura2008magnetically}.

However, the true structure of magnetic fields in molecular clouds is certainly more complex. If we consider the entire Taurus cloud, rather than just B211/3, a wider range of relative orientations between filaments and fields is seen \citep{goodman1990optical}. For example, B18 is one of the denser filaments in Taurus and appears to have a coherent magnetic field with an orientation that differs by $\approx45^\circ$ from the filament orientation. While this is not necessarily inconsistent with collapse along field lines it does imply that this more complex cloud requires similarly more complex simulations. More importantly, progress in this area requires sensitive magnetic field observations for a statistically significant sample of molecular clouds. Upcoming submm polarimetry results from the \blast{} 2012 flight and future \blast{} flights as well as from \textit{Planck} \citep{2013arXiv1303.5062P} should be especially helpful in this regard.

\section{Summary}

We have presented degree-scale \three{} and \five{} polarization maps of the Lupus I molecular cloud, obtained from the 2010 flight of \blast{}. Despite the systematic error introduced by the damaged spectral filter, our data provide new information on the magnetic field in this cloud. We see a reasonably close alignment between the primary filament normal and the magnetic field direction as measured by both optical and submm polarimetry. Specifically, the submm and optical results from the Uniform Field model differ by $\approx20^\circ$, and the results from the Constant Offset model show that the submm and optical field directions bracket the filament normal, differing from it by $9.8^\circ$ and $8.6^\circ$, respectively. The B228 outflow position angle and the submm Uniform Field angle differ by $\approx30^\circ$, but without constraints on the inclination angle this value cannot be directly compared with other results. The Lupus I cloud appears to be consistent with the general picture of a primary filament approximately perpendicular to the large-scale field, with secondary filaments running nearly parallel to the field, as is also seen in Serpens South \citep{sugitani2011near} and in the  model of \citet{nakamura2008magnetically}.

The \blast{} collaboration acknowledges support from NASA (through grant numbers NAG5-12785, NAG5-13301, NNGO-6GI11G,  NNX0-9AB98G, and the Illinois Space Grant Consortium), the Canadian Space Agency (CSA), the Leverhulme Trust through the Research Project Grant F/00 407/BN, Canada's Natural Sciences and Engineering Research Council (NSERC), the Canada Foundation for Innovation, the Ontario Innovation Trust, the Puerto Rico Space Grant Consortium, the Fondo Institucional para la Investigaci{\'o}n of the University of Puerto Rico, and the National Science Foundation Office of Polar Programs. C. B. Netterfield also acknowledges support from the Canadian Institute for Advanced Research. Finally, we thank the Columbia Scientific Balloon Facility (CSBF) staff for their outstanding work.

\nocite{basu2000magnetic}

\bibliography{ref}

\appendix

\section{Appendix 1} \citet{rizzo1998starlight} presented optical polarimetry on two populations of stars, which they refer to as Field Stars and Weak Stars. All Field Stars have HD catalog numbers, and their R.A. and Decl. coordinates were retrieved from the catalog. Polarization data for the Weak Stars were presented in Table 1 of \citet{rizzo1998starlight}, but sky locations were presented only graphically (in a finding chart). The finding chart clearly shows many optical stars and indicates the sky location of the Weak Stars on the high extinction region of the Lupus I filament. To obtain sky coordinates for the Weak Stars, a DSS2 map of the region obtained from Skyview\footnote{http://skyview.gsfc.nasa.gov/} was superimposed on the finding chart. The sky coordinates of reference stars in the finding chart were obtained and the coordinates of the Weak Stars were triangulated from them. Reference stars were added until the positions of the weak stars stabilized to within the uncertainty in their location on the finding chart, which was $\approx10$-$20''$. An updated version of Table 1 from \citet{rizzo1998starlight} is presented in Table \ref{tab:optical} containing the derived Weak Star R.A. and Decl. coordinates.

\begin{deluxetable}{lccc}
\tablewidth{0pt}
\tablecaption{\label{tab:results}Fits to Magnetic Field Models}

\tablehead{\colhead{} & \colhead{} & \multicolumn{2}{c}{Model Type} \\
\cline{3-4}
\colhead{} & \colhead{No.\ of} & \colhead{Uniform Field} &
\colhead{Constant Offset\tablenotemark{a}} \\
\colhead{Data} & \colhead{pseudo-vectors} & \colhead{(deg)} & \colhead{(deg)}}

\startdata
All submm    & 56 & 34.3 & \phantom{1}$-$9.8 \\
$350\:\mu$m  & 26 & 28.9 & $-$15.1 \\
$500\:\mu$m  & 30 & 39.2 & \phantom{1}$-$5.3 \\
All optical  & 38 & 53.9 &  \phantom{1}8.6 \\
Near optical\tablenotemark{b} & 20 & 53.1 & \phantom{1}9.0 \\
Corrected optical\tablenotemark{c} & 38 & 50.1 & \phantom{1}5.1 \\
\enddata

\tablenotetext{a}{Difference between the inferred field direction at a given point and a model field oriented at a constant angle relative to the normal to the blue arc shown in Figure \ref{fig:avmap} (see section \ref{sec:magnectic_models}).}
\tablenotetext{b}{Optical pseudo-vectors within $15'$ of a submm pseudo-vectors.}
\tablenotetext{c}{Optical pseudo-vectors after subtracting contribution from 
Field Stars.  See Section \ref{sec:compare_optical} for details.}
\end{deluxetable}
 \begin{deluxetable}{llccrr}
\tablewidth{0pt}
\tablecaption{\label{tab:optical}Optical Pseudo-vectors for Weak Stars\tablenotemark{a}}
\tablehead{\colhead{R.A.} & \colhead{Decl.} & \colhead{$P$} & \colhead{$\sigma_P$} &
\colhead{$\theta$\tablenotemark{b}} & \colhead{$\sigma_\theta$} \\
\colhead{(J2000)} & \colhead{(J2000)} & \colhead{(\%)} & \colhead{(\%)} &
\colhead{(deg)} & \colhead{(deg)}}

\startdata
234.53204 &	$-$33.253026 & 2.50 & 0.36 &  36.4 &  4.1 \\
234.74343 &	$-$33.271377 & 2.60 & 0.20 &  42.5 &  2.2 \\
234.43411 &	$-$33.404653 & 1.88 & 0.53 &  53.2 &  8.1 \\
234.85885 &	$-$33.412643 & 0.72 & 0.15 & $-$10.0 &  5.9 \\
235.06775 &	$-$33.545955 & 1.35 & 0.10 & 108.8 &  2.2 \\
234.70345 &	$-$33.586962 & 2.11 & 0.25 &  44.2 &  3.4 \\
234.90323 &	$-$33.669157 & 1.71 & 0.42 &  91.0 &  7.0 \\
235.05416 &	$-$33.723124 & 1.62 & 0.23 &  27.6 &  4.1 \\
235.20059 &	$-$33.668061 & 0.71 & 0.10 &  58.9 &  4.0 \\
235.23064 &	$-$33.826070 & 1.43 & 0.27 &  42.7 &  5.3 \\
235.33343 &	$-$33.770209 & 0.84 & 0.13 &  57.1 &  4.3 \\
235.60495 &	$-$33.753452 & 1.24 & 0.20 &   4.9 &  4.5 \\
235.44008 &	$-$33.975295 & 2.19 & 0.29 &  52.1 &  3.8 \\
235.58651 &	$-$34.016795 & 1.80 & 0.13 &  63.2 &  2.1 \\
235.61949 &	$-$33.859589 & 1.34 & 0.23 &  36.1 &  4.9 \\
235.68931 &	$-$33.846024 & 0.85 & 0.34 &  37.0 & 11.5 \\
235.80531 &	$-$33.976287 & 0.81 & 0.08 &  62.3 &  2.9 \\
235.90906 &	$-$34.078436 & 1.24 & 0.19 &  37.0 &  4.5 \\
235.48180 &	$-$34.226688 & 0.90 & 0.10 &  47.3 &  3.2 \\
235.58846 &	$-$34.219505 & 0.95 & 0.17 &  50.7 &  5.2 \\
235.81402 &	$-$34.262761 & 2.54 & 0.27 &  72.1 &  3.1 \\
235.95366 &	$-$34.163824 & 2.00 & 0.35 &  32.1 &  5.0 \\
236.04189 &	$-$34.256393 & 2.46 & 0.16 &  34.3 &  1.9 \\
236.00137 &	$-$34.379713 & 2.08 & 0.18 &  70.5 &  2.5 \\
236.02853 &	$-$34.555294 & 0.93 & 0.13 &  77.5 &  3.9 \\
236.02077 &	$-$34.683778 & 1.07 & 0.11 &  73.3 &  3.0 \\
236.48068 &	$-$34.293946 & 0.96 & 0.14 &  12.1 &  4.1 \\
236.49813 &	$-$34.446355 & 1.99 & 0.44 &  81.6 &  6.3 \\
236.66298 &	$-$34.475896 & 1.08 & 0.14 &  46.9 &  3.6 \\
236.67918 &	$-$34.595195 & 1.31 & 0.20 &  56.8 &  4.3 \\
236.41930 &	$-$34.612734 & 2.47 & 0.19 &  71.1 &  2.2 \\
236.62312 &	$-$34.700913 & 0.70 & 0.13 &  59.2 &  5.5 \\
236.82965 &	$-$34.772735 & 1.41 & 0.14 &  34.6 &  2.8 \\
236.90333 &	$-$34.915581 & 2.23 & 0.14 &  30.5 &  1.8 \\
236.90432 &	$-$35.101516 & 1.80 & 0.16 &  80.1 &  2.6 \\
236.67547 &	$-$35.081575 & 1.15 & 0.11 &  58.9 &  2.7 \\
237.02646 &	$-$35.180664 & 2.15 & 0.33 &  77.5 &  4.4 \\
236.50093 &	$-$35.327369 & 1.29 & 0.07 &  70.2 &  2.2 \\
236.56392 &	$-$35.466378 & 1.11 & 0.51 &  51.1 & 13.2 \\
236.78695 &	$-$35.412109 & 1.19 & 0.14 &  70.3 &  3.5 \\
\enddata

\tablenotetext{a}{From \citet{rizzo1998starlight}. See Appendix 1 for details.}
\tablenotetext{b}{Position angle of the polarization $E$-vector, measured
east of north.}
\end{deluxetable}

\end{document}